# Direct Visualisation of Out-of-Equilibrium Structural Transformations in Atomically-Thin Chalcogenides

**Pawan Kumar**[1,2]**, James P. Horwath**[2]**, Alexandre C. Foucher**[2]**, Christopher C. Price**[2]**, Natalia Acero**[3]**, Vivek B. Shenoy**[2]**, Eric A. Stach**[2#] **and Deep Jariwala**[1#]**.**

[1]Department of Electrical and Systems Engineering, University of Pennsylvania, Philadelphia-19104, USA.

[2]Department of Materials Science and Engineering, University of Pennsylvania, Philadelphia-19104, USA.

[3]Vagelos Institute for Energy Science and Technology, University of Pennsylvania, Philadelphia-19104, USA.

**#E-mail: dmj@seas.upenn.edu; stach@seas.upenn.edu**

## Abstract:

Two-dimensional (2D) transition metal dichalcogenides (TMDCs) have been the subject of sustained research interest due to their extraordinary electronic and optical properties. They also exhibit a wide range of structural phases because of the different orientations that the atoms can have within a single layer, or due to the ways that different layers can stack. Here we report the first study of direct-visualization of structural transformations in atomically-thin layers under highly non-equilibrium thermodynamic conditions. We probe these transformations at the atomic scale using real-time, aberration corrected scanning transmission electron microscopy and observe strong dependence of the resulting structures and phases on both heating rate and temperature. A fast heating rate (25 °C/sec) yields highly ordered crystalline hexagonal islands of sizes of less than 20 nm which are composed of a mixture of 2H and 3R phases. However, a slow heating rate (25 °C/min) yields nanocrystalline and sub-stoichiometric amorphous regions. These differences are explained by different rates of sulfur evaporation and redeposition. The use of non-equilibrium heating rates to achieve highly crystalline and quantum-confined features from 2D atomic layers present a new route to synthesize atomically-thin, laterally confined nanostrucutres and opens new avenues for investigating fundamental electronic phenomena in confined dimensions.

**Keywords:** two-dimensional, quantum dots, $MoS_2$, phase transformation, heterojunctions, electron microscopy.

## Introduction:

Two-dimensional nanomaterials have received significant attention due the wide range of exciting properties that they exhibit[1]. Transition metal dichalcogenides (TMDCs) have garnered particular interest due to their ability to be utilized in novel electronic device applications[2]. There have been several studies investigating the effect of heat treatments on the structural properties of these systems, but to date these have largely focused on amorphous to crystalline transformation kinetics, point defect kinetics, and dislocation kinetics under equilibrium conditions from ranging ambient to elevated temperatures[3, 4]. Among 2D materials, the transition metal dichalcogenides (TMDCs) of Mo and W can exist in multiple phases induced by atomic distortions, including the 2H, 1T and 1T' phases. Furthermore, additional phases, namely 2H and 3R, are formed by the stacking of 2D layers with different rotaitional arrangements. In particular, $MoS_2$ – both as a monolayer and in multilayer form – is known to exhibit multiple structural phases. These include the chalcogen distortion-induced 1T phases for both monolayers and thicker crystals, and stacking alignment-induced 3R phases in multilayers[5-8]. The 2H phase is the most thermodynamically stable phase, while the 1T phase and 3R phases are metastable and revert back to the 2H phase over time, or upon application of some form of energy that allows them to overcome the associated activation barrier[6, 9]. Despite the fact that several studies regarding phase identification and evolution have been conducted, the nucleation and formation of metastable phases is still a topic of major debate and discussion[10-14]. Thus far, in-situ electron microscopy studies of TMDCs (such as $MoS_2$, $WS_2$, etc.) have largely focused on observations of the kinetics of individual point defects[15, 16]. However, with the rapid growth of interest in the optical, electronic, and mechanical properties of TMDCs[17, 18], several studies have used *in situ* scanning transmission electron microscopy (STEM) to document structural transformations under equilibrium conditions[4, 19-21]. In addition, some observations of in-situ heterojunction formation and growth have also been made in the same class of materials[22, 23]. These nanoscopic investigations have shown that defects and vacancies form due to the application of multiple types of external stimuli, along with the dynamic migration of transition metal and chalcogen atoms towards favourable grain boundary formations[24]. There is, however, a lack of understanding on how processing conditions impact atomic scale structure and phase evolution in confined dimensions, including how they lead to the formation of new, meta-stable phases. [6, 7, 25]

Here, we have studied the effect of equilibrium vs. non-equilibrium diffusion conditions during the heating of atomically thin 2D $MoS_2$ sheets. We find that the effect of heating rate on the morphology of bilayer/few-layer $MoS_2$ is consistent with the equilibrium bulk binary phase diagram of Mo and S[26, 27]. We see that stoichiometric $MoS_2$ can phase-separate at low temperatures (T < 500 ˚C) into a two-phase region of Mo + $MoS_2$ if the sulfur concentration is reduced. At higher temperatures (500 < T < 1000 ˚C), the same process leads to a mixture of $Mo_2S_3$ and 3R phase of $MoS_2$. Furthermore, if the sulfur content is reduced to below 60 atomic percent, a mixture of Mo and $Mo_2S_3$ forms. It has been established in the literature that sulfur vacancies, having a low formation energy, are the dominant defect species in few-layer $MoS_2$[28], and that annealing in the 500 ˚C < T < 1000 ˚C range can lead to loss of sulfur atoms on a time scale of just 30 minutes[29]. Moreover, the general phenomena of melting point depression in quantum-confined materials relative to their bulk counterparts is well known[30, 31]. Based on these concepts, we propose that changing the heating rate of a bilayer $MoS_2$ sample can be used as a proxy to tune the sulfur concentration and change the material structure and morphology of a stoichiometric crystal after synthesis. Direct observation of atomic level diffusion indicates the different crystalline states across the entire thermal diffusion process. Lateral heterophase formation along with statistical observation of particle evolution provides a new pathway for materials engineering in the quantum mechanically-relevant size regime[32].

## Results and Discussion:

### Phase transformation processes in equilibrium vs non-equilibrium heating conditions:

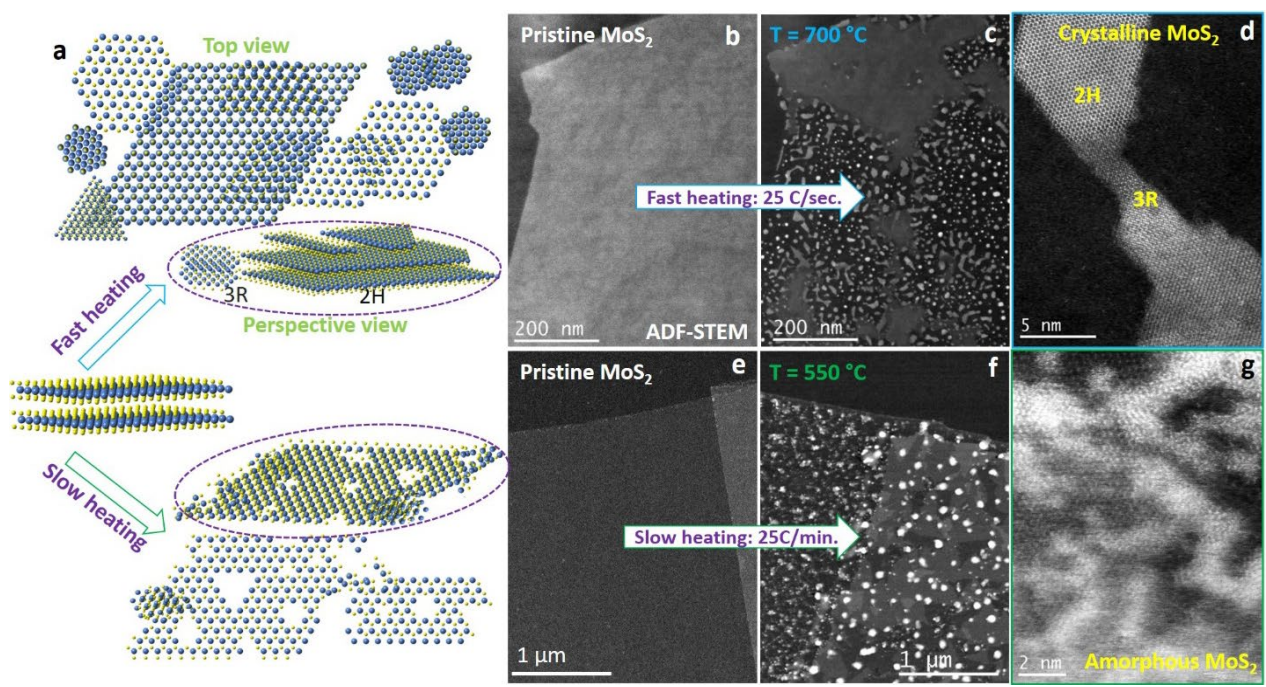

**Figure 1: Direct visualisation of diffusion processes for non-equilibrium vs equilibrium environments.** (a) Schematic shows the diffusion pathways for disintegration vs. decomposition of 2D layers in two discrete environments, (b, e) STEM imaging of pristine bilayer $MoS_2$, as transferred to the MEMS chip and supporting $SiN_x$ membrane grid and corresponding (c, f) disintegrated crystalline nano-flakes and decomposed $MoS_2$ respectively, (d) Disintegrated crystalline $MoS_2$ contains multiple phases while (g) equilibrium decomposed $MoS_2$ into amorphous / highly disordered structures. Atomic scale STEM images in Figure d and g taken from the corresponding section of Figure c and f respectively.

Schematics presented in Figure 1a show the thermal diffusion process in 2D layered $MoS_2$ crystals, mechanically exfoliated from bulk crystals, when subjected to two different heat treatments: 1) localised, fast heating rates, using embedded heating elements on a chip, which create highly non-equilibrium thermodynamic conditions, and 2) global, slow heating in an equilibrium environment in a hot-walled reactor. To accomplish these experiments and visualize their atomic structure, exfoliated $MoS_2$ layers were transferred to heater-embedded TEM grids (Hummingbird Scientific)[33]. HAADF-STEM images in Figures 1b and 1e represent the as-prepared 2D $MoS_2$ layer before thermal treatment, while Figures 1c and 1f show the transformed $MoS_2$ after thermal treatment in non-equilibrium and equilibrium conditions, respectively. Similar results for another set of samples are shown in Figure S1 (supporting information) to confirm the reproducibility of our experiments. In the case of non-equilibrium (fast) heating, we observe discrete integration of continuous, atomically thin and uniform layers of $MoS_2$ into highly crystalline nanostructures or nanoparticles with lateral dimensions of less than 20 nm, which is within the regime of electronic quantum confinement. Figure 1d presents examples of nanoscale islands of $MoS_2$ that form after fast ramping of the flakes shown in Figure 1b to elevated temperature (700 ˚C). Conversely, a slow increase in temperature results in decomposition of the stoichiometric, single crystalline $MoS_2$ layers into amorphous structures. The sections below discuss the detailed mechanisms and structures observed in each case and correlate them with theoretical insights which account for different evaporation rates of sulfur from the structures.

**Macroscopic thermal diffusion in non-equilibrium conditions and the restructuring of heterophase formation:**

While studies of phase evolution through heating or beam induced transformation have been previously carried out on 2D layered chalcogenides, those studies have been limited to small area, atomic scale investigations which offer limited insight into dynamic phenomena. Large area analysis and observations at different length scales using electron microscopy has been lacking. For the case of fast heating, we show this in Figure 2a-b where a magnification series of images shows how the disintegrated portion of a $MoS_2$ flake taken through a rapid heating cycle develops, including analysis of its atomic structure and crystalline order. It is clear from these images that upon rapid heating the $MoS_2$ single crystal flake disintegrates into highly crystalline, nanosized islands that retain hexagonal symmetry. This is also evident from the selected area electron diffraction (SAED) patterns in Figures 2a-b.

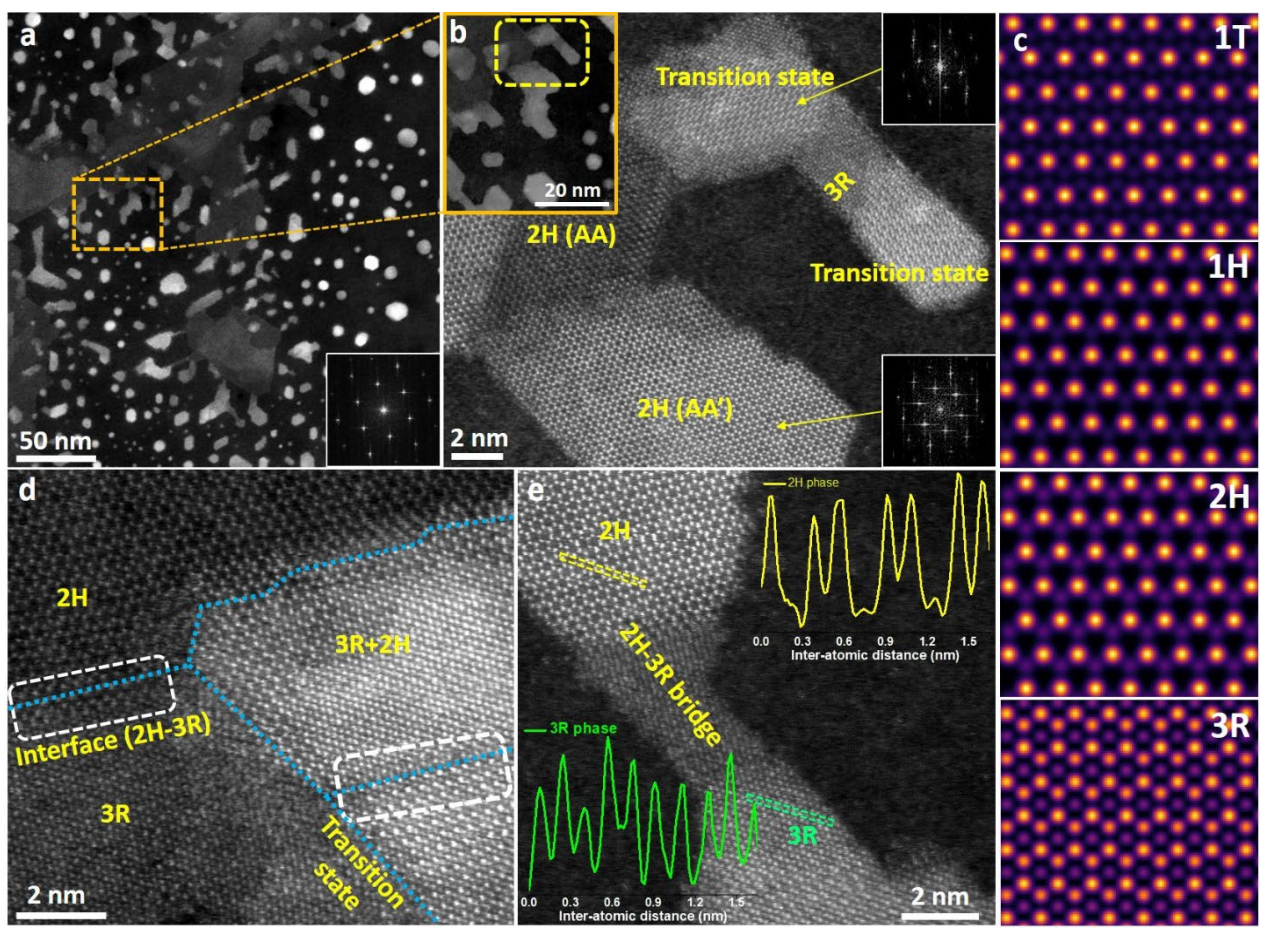

**Figure 2: Crystalline disintegrated MoS₂ nanoflakes formed after in-situ fast heating.** (a) Low magnification ADF STEM image and inset selected area electron diffraction (SAED) pattern shows the high degree of crystallinity of disintegrated MoS₂ from the large 2D sheets, (b) Rotated, high magnification STEM image of highlighted boxed region in Figure 2-a of MoS₂ sheet after disintegration. The atomic scale HAADF STEM image shows the atomic arrangement for differently stacked MoS₂. (c) STEM image simulation performed for different poly-phases of MoS₂ to correlate with the experimentally observed atomic level images. (d) Atomically resolved STEM image shows formation of a heterojunction between the 2H and 3R phases, a mixed state layer (2H+3R) along with a transition state of restructured 2H MoS₂. (e) Line profile across the two different phases, 2H and 3R present in the disintegrated nanoflake of MoS₂ shows the atomic placement in the corresponding structures.

Additional atomic resolution STEM images and compositional analysis via energy dispersive X-ray spectroscopy (EDS) are provided in Supporting Figures S2 and S3. In addition to observing the highly ordered nature of the disintegrated MoS₂ crystal, we also observe polymorphic phase formation, as highlighted in Figure 2b and 2d. SAED and EDS analysis indicated that the crystal symmetry of the 2H phase is intact. Figure 2b clearly shows the formation of different phases and structural rearrangements of MoS₂ nanostructures, determined to be the 2H and 3R phases. In Figure 2c, we present image simulations (QSTEM package)[34] using the same experimental parameters to replicate the observed contrast and structural arrangements seen in the experimental HAADF STEM images. Only the 2H and 3R phases match with the experimentally observed contrast, and we do not see formation of any other phases, such as 1T and 1T', elsewhere. However, we did find regions interconnecting the 2H and 3R phases which appear as transition states (Figure 2b, d). Additional analysis of structural arrangement of atoms and stacking order of planes for different phases in these transition regions is provided in Figure 2d and 2e. Atomically sharp heterojunction formation is observed at the intersection of the 2H and 3R regions (Figure 2d) in partially disintegrated regions, suggesting that the equilibrium 2H phase transforms into a metastable 3R phase in certain regions following a fast-thermal treatment. We speculate that the heat supply and dissipation across the flake is non-uniform and the partially disintegrated regions occur when the top atomic layers of the crystal have already disintegrated but the bottom layers have yet to disintegrate at the time of termination of the experiment. We observe that the occurrence of the 3R phase is strictly limited to fully/partially disintegrated regions of the original 2H-phase continuous crystal. The continuous regions of the original crystal that do not disintegrate all remain in the pure 2H phase, while the partially disintegrated islands show a mixture of 2H and 3R phases (Figure S5). We hypothesize that the 3R phase is formed and stabilized in this case as result of restacking and vertical mass transport of the van-der-Waals layers during fast thermolysis of $MoS_2$[9]. Atomic scale line profiles along $[11\overline{2}0]$ of the two phases have also been plotted in the inset of Figure 2e. This clearly resolves the atomic locations (peak placement) and chemistry (peak height, equivalent to $\sim Z^2$ contrast) in the two different structures. Additional images and analysis related to the structural model, distributions of 2H and 3R phases, as well as thickness variation after disintegration are provided in supporting information Figures S4, S5, S6 and Video V1.

**Coarsening and Epitaxial alignment:**

The above discussion describes the structure of the disintegrated phase upon rapid heating. While this provides new insight into the structural evolution, the dynamics of the transformation are difficult to capture in real-time measurements. This is particularly challenging since, upon reaching the desired temperature, layer disintegration is very rapid and occurs over the acquisition time of only two or three STEM images. We have therefore collected a sequence of bright field TEM images at different times (Figure 3), at a fixed temperature (650 ˚C) using a high-frame rate camera (GATAN One-View IS)[35]. The sample was slowly heated from room temperature to 550 ˚C and then rapidly heated from 550 ˚C to 650 ˚C and observed *in situ*.

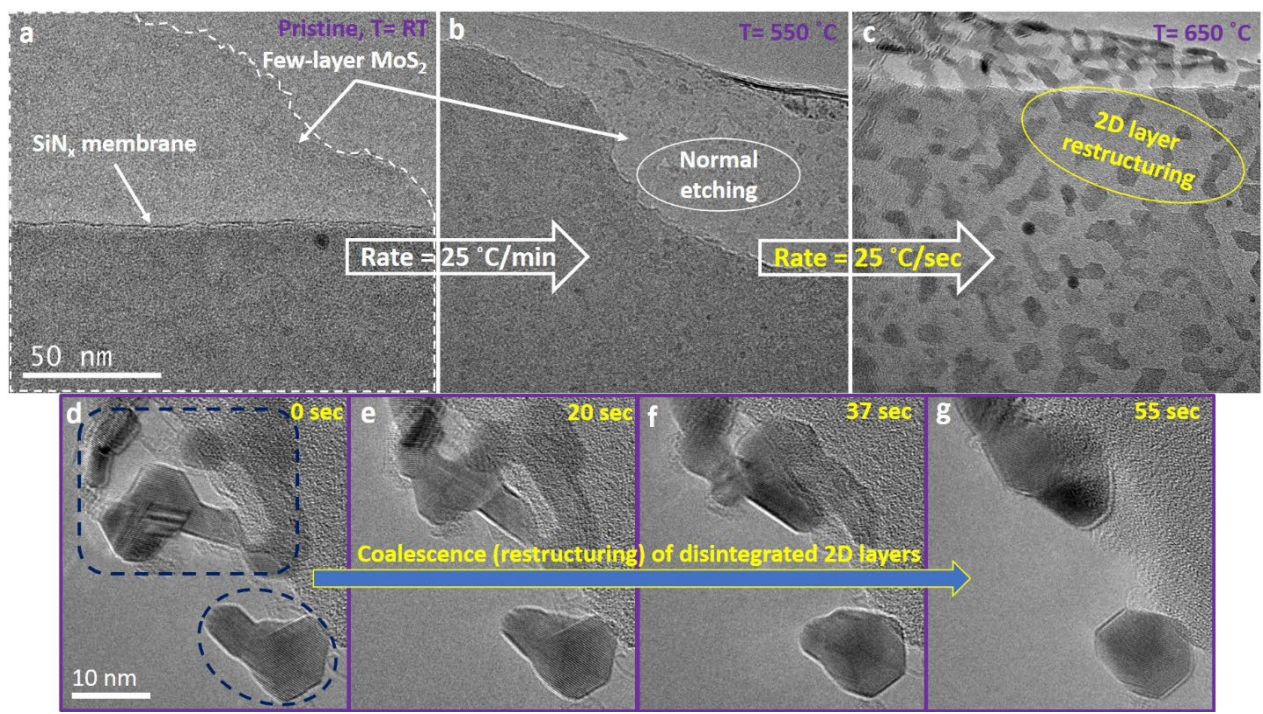

**Figure 3: Real time observation of few-layer MoS₂ disintegration and subsequent coarsening using BF-TEM imaging.** (a) Pristine few-layer MoS₂ at room temperature and the corresponding (b) formation of small etched regions when slowly (~25 ˚C/min) heated to 550 ˚C. (c) Complete disintegration of same few layer MoS₂ when ramped to 650 ˚C with fast heating rate (~25 ˚C/sec), (d-g) Sequential snapshots of BF-TEM images taken at different times showing further coarsening of the disintegrated MoS₂.

Figures 3a and 3b show representative images of few-layer MoS₂ when heated to 550 ˚C at a slow ramping rate (25 ˚C/min). Small etch pits form, (Figure 3b) which suggests slow evaporation of chalcogen atoms and associated defect creation in the MoS₂ lattice, see Figure S7 for more details. Upon ramping at an accelerated rate of 25 ˚C/sec from 550 ˚C to 650 ˚C, we observe a dramatic transformation of the MoS₂ sheet into disintegrated nanoscopic domains or islands (Figure 3c and Figure S8, supporting information). We have also examined the non-irradiated regions (i.e. those not exposed to the electron beam) and find that the MoS₂ sheet undergoes a similar disintegration process after reaching 650 ˚C. Further analysis of sequential phase contrast TEM images (Figure 3d-g), shows the different stages of coarsening of the MoS₂ nanostructure when held at a constant elevated temperature. It is worth noting that with rapid heating these nanostructures spontaneously form on a time scale that is less than the 100 fps frame rate of the camera. In contrast, their slow evolution when held at a constant temperature can be captured. As seen in Figure 3d-g, after 55 seconds, multiple smaller crystallites of $MoS_2$ have coalesced into a larger crystal of ~10 nm in diameter with a hexagonal shape. For Figure 3d-g, the coalescence is prominently visible in the bottom right corner of the image (blue dashed oval). We speculate that the initial disintegration mechanism breaks down the continuous few-layer thick $MoS_2$ into regions with geometrical shapes based on energetically favourable facets. As the sample is held at an elevated temperature over time, partial etching ensues resulting in the loss of sulfur atoms via evaporation and the migration of atoms overall. With time, newly formed islands further restructure to reduce their surface energy leading to strongly faceted particles. Figures 3 d-g show that the island extends only slightly in the lateral dimension as the shape evolves, and that the thickness increases. This is evident from the decreased prominence of lattice fringes and darker contrast in the centre of the island. These observations are consistent with well-established theories of particle coalescence, wherein surface diffusion at the point of contact drives restructuring to reduce the total local curvature[36].

It is noteworthy that this thermally driven spontaneous disintegration into nanoscale islands occurs in both free standing as well as substrate supported crystals, as seen in Figure 3c. While there appears to be a partially connected network of particles in the free-standing region, on the amorphous non-epitaxial SiNx membrane support the particles are isolated and separated into "quantum dot" sized islands (See Figure S6 in supporting information for more images and video V2 for dynamics of coarsening). The high degree of crystallinity of the lattice both during coarsening (see video V2 in supporting information) and in the sub 10 nm size islands is remarkable and has not been seen before via chemical exfoliation or bottom up synthetic routes for Mo and W chalcogenides, to our knowledge.

Epitaxy and lattice matching are not often discussed in the context of layered 2D material growth and synthesis, primarily because weak van der Waals (vdW) interlayer interactions generally preclude any significant influence by the substrates on subsequent layer orientation and growth. However, we see here that the van der Waals interactions, though weak in nature, can influence growth and epitaxial alignment of the nanoscale islands. Similar van der Waals epitaxy has been exploited for aligned growth of heterostructures on 2D layered materials, as a well as TMDC growth on standard substrates[37-43].

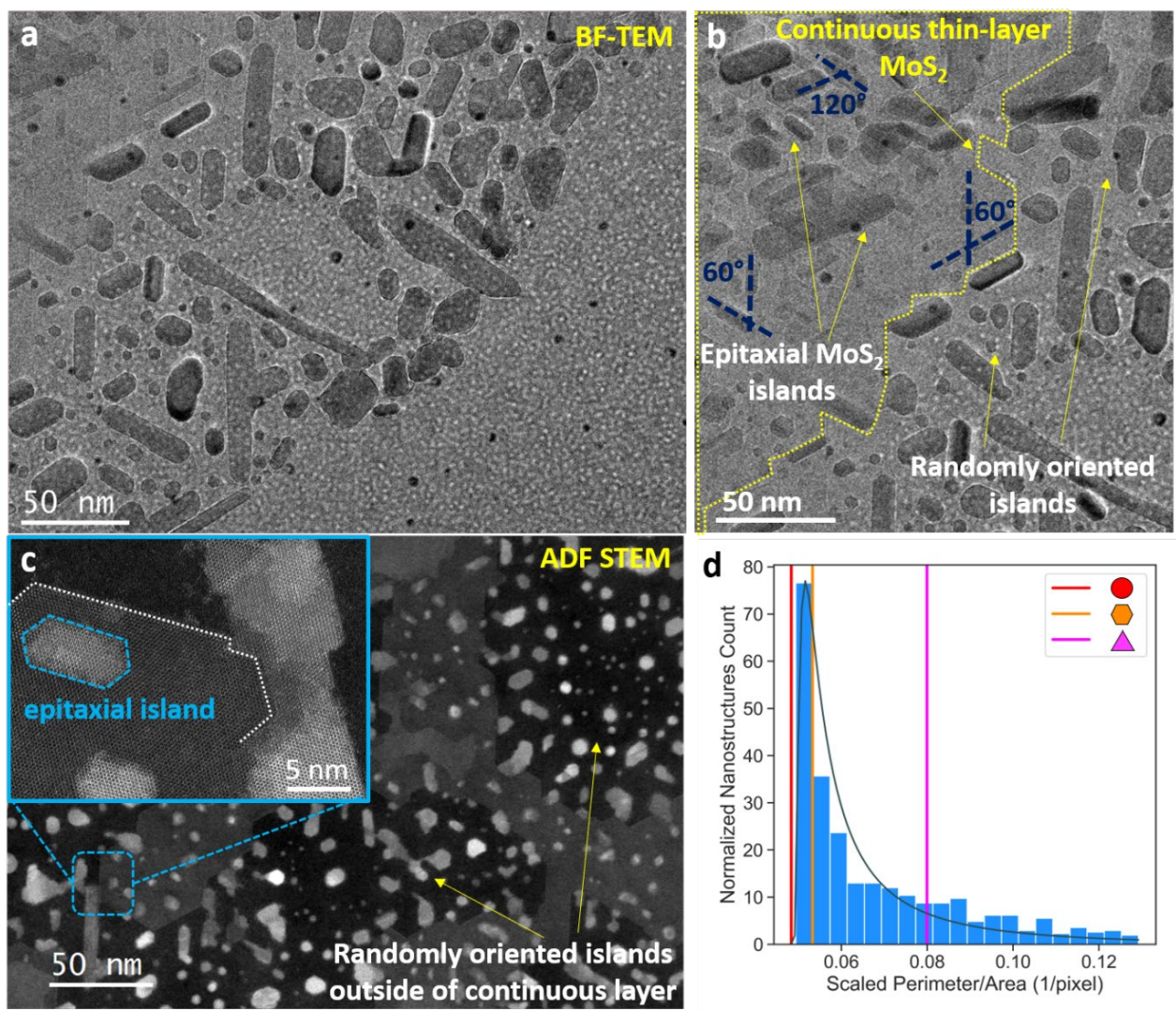

**Figure 4: Epitaxy and orientation during the disintegration of few-layer MoS₂ on in-situ MEMS chip.**
(a) Large area bright field (BF)-TEM image of the disintegrated portion of MoS₂ where a portion of the flake has completely disintegrated. (b) Partially disintegrated MoS₂ with a supporting, continuous layer of MoS₂ underlying the islands, leading to epitaxial alignment of the islands over the continuous single crystalline layer. (c) Continuous bilayer MoS₂ disintegrated at 700 ˚C upon rapid heating showing islands on top of continuous regions as well as fully disintegrated regions. Inset presents an atomic resolution image demonstrating epitaxial alignment in the partially decomposed region and randomly oriented islands in the completely disintegrated region. (d) Quantitative measurement of island size and shape after the non-equilibrium diffusion. Legend details described in main text.

Furthermore, upon rapid heating, we observe an interesting epitaxial relationship between fully and partially disintegrated regions of few-layer MoS₂ deposited on an amorphous support layer (SiNx), which is detailed in Figure 4. High magnification TEM images show that the few-layer MoS₂ goes through a layer-by-layer disintegration, followed by coarsening of the 2D layers. Figure 4a presents a low magnification TEM image that shows that a large portion of the MoS₂ layer has fully disintegrated into faceted nanostructures, while a small portion in top-left corner remains only partially disintegrated. Analysis of the geometric alignment of the nanoscale $MoS_2$ islands indicates that the $MoS_2$ nanostructures in the fully disintegrated region do not exhibit any preferential alignment or epitaxy with one another. However, the islands that sit above a continuous single crystalline $MoS_2$ layer exhibit some preferential alignment (Figure 4b). We observe that the fully disintegrated $MoS_2$ nanostructures are transformed into "quantum dot" or faceted nanostructures with random orientations and alignment with respect to one another. In contrast, in the partially disintegrated regions (Figure 4c inset) the nanoscale islands maintain perfect epitaxial orientation with the remaining continuous $MoS_2$ layer. Since the sample thickness is much smaller than the perpendicular phonon mean free path[44], we propose that the disintegration mechanism occurs via point defect formation and aggregation[16, 45] (Figure 6). To minimize disruption of vdW interactions, these point defects will preferentially form in the layer furthest from the substrate, leaving the layer adjacent to the substrate intact for longer time and leading to the observed epitaxy. Additional characterization details of the nanoscale islands including information regarding their composition is provided in Figure S10.

Our large area, in-situ analysis allows for shape characterization of the population of nanocrystalline islands which form after rapid heating and annealing at 650 ˚C. The ADF STEM images such as Figure 4c are segmented using a connected components algorithm[46] to identify each nanoscopic island of $MoS_2$ from the void or substrate background. The perimeter and area of each island are then measured in terms of number of pixels on the edge and the interior of the island, respectively. We then scale the perimeter of each island to the average (260 pixels) and scale the area by the square of the scale factor. This enables us to use the perimeter to area ratio to characterize the shape of each island, removing the effect of dilation. A histogram of this shape factor is plotted in Figure 4d for the smallest 775 nanostructures, comprising 90% of the islands detected by the connected components algorithm. The lines overlaid for reference correspond to the shape factors of regular polygons with equivalent perimeters, with the circle having the smallest perimeter / area ratio. The regular hexagon and equilateral triangle are also shown since these shapes have been commonly observed in CVD growth experiments of TMDCs[47]. We see that the majority of the nanoscale islands adopt a shape somewhere between the circle and hexagon, suggesting a rounded hexagonal morphology, broadly consistent with the HAADF-STEM images in Figure 4. The distribution of islands can be empirically fit by an inverse Gaussian $f(x, \mu) = \frac{1}{\sqrt{2\pi x^3}} \exp(-\frac{(x-\mu)^2}{2x\mu^2})$ with a shape factor $\mu$ of 2.1. A non-normal distribution like this is expected due to the energetic preference for specific edge configurations[48]. The dominance of the hexagonal shape is indicative of a neutral chemical potential environment with a stoichiometric balance and a mixture of chalcogen and metal-terminated edges. Triangular flakes tend to form in extremums of the chemical potential since the edges are either all chalcogen or have a metal-rich zig-zag structure[49]. This is further evidence that the rapid in-situ heating rate preserves the composition of the lattice at the initial MoS$_2$ ratio. This will be discussed in more detail below (see Figure 6).

**Decomposition and amorphization of MoS$_2$ under slow, equilibrium heating:**

The above sections discuss the structural changes that occur during the rapid heating of 2D MoS$_2$. To isolate the specific effects of the heating rate, we performed a comparative analysis with ex-situ, slow heating of the MoS$_2$ layers on SiNx membranes. A few prior studies have suggested that slow formation of vacancies due to chalcogen removal and migration at elevated temperatures that ultimately leads to the formation of voids. Our observations agree with these earlier reports in monolayer to few-layer MoS$_2$.

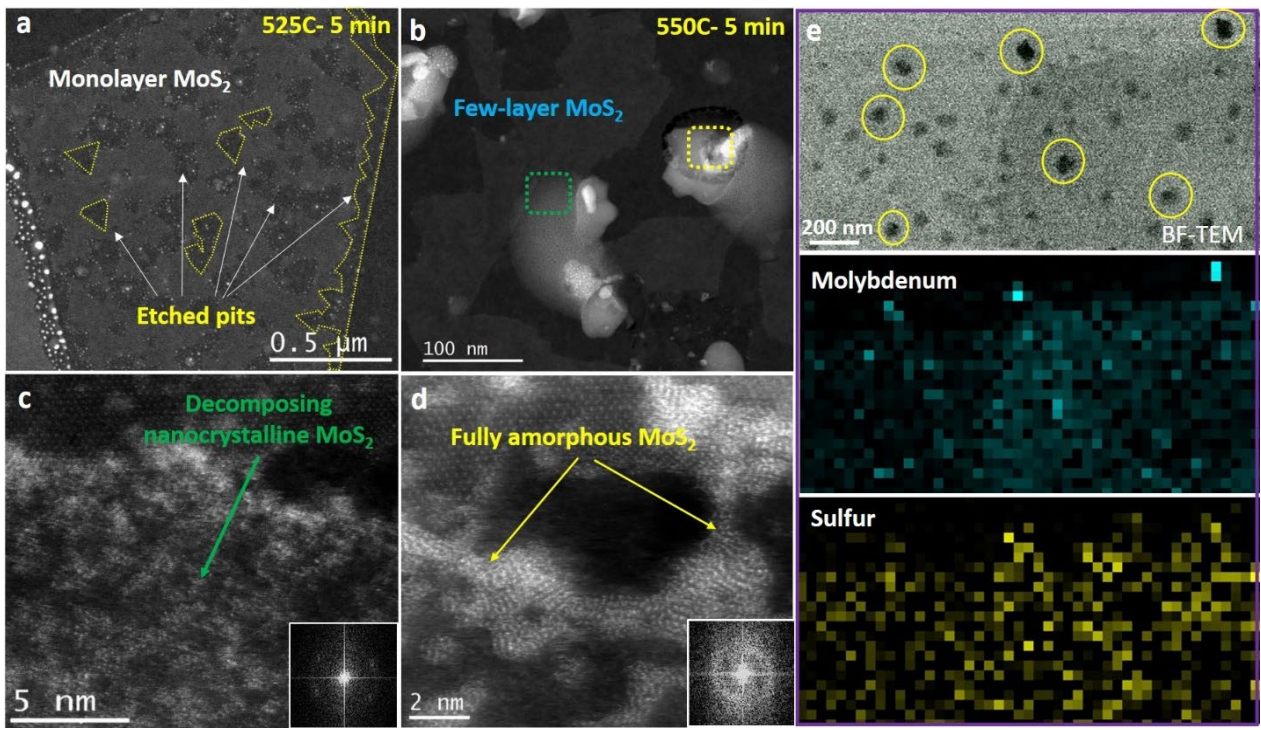

**Figure 5: Ex-situ heated (rate = 25 ºC/min) MoS$_2$ layers in equilibrium conditions.** (a) HAADF-STEM imaging of the MoS$_2$ heated to 525 ˚C for 5 minutes. Triangular etch pits inside the monolayer along-with completely etched edges are evident. (b) Images from a few-layer MoS$_2$ region heated in these same conditions, and at higher magnification. (c) HAADF-STEM image of few layer MoS$_2$ heated to 550 ˚C for 5 minutes. Nanocrystalline order is evident in corresponding FFT which shows faint, diffused spots (inset) (d) Fully decomposed amorphous MoS$_2$ with along with corresponding FFT inset (550 ˚C for 5-10 mins). FFT patterns (inset in c and d) represent the crystallinity of the corresponding region indicated by the green and yellow dashed rectangular regions marked in Figure 5 (b). (e) BF-STEM image and corresponding EDX maps of molybdenum (cyan) and sulfur, (yellow). The yellow circles indicate regions of nanocrystal formation which are highly enriched in Mo, ultimately leading to formation of pure Mo metal crystals.

Figure 5 presents a detailed analysis for describing these phenomena over a wide range of temperatures and over a large area. Ex-situ thermal treatments were carried out in a quartz tube furnace in two different environments (vacuum as well as pure Ar gas) using the slow heating rate (25 ˚C/min), as detailed in the methods section. The slow heating rate and the uniformity of temperature inside the tube furnace makes this an equilibrium heating condition, leading to a fundamentally different disintegration process (called decomposition) than the one observed in the non-equilibrium (rapid heating) case. Slow heating to reach elevated temperatures results in formation and migration of vacancies, starting with the ones that have the lowest formation energies as described in the literature [24, 50, 51]. This results in the formation of nanoscale voids in the layers, as shown in Figure 5a and 5b. Another distinctive feature of slow-heating induced disintegration is that the affected region is localized to high energy or defective regions such as the flake edges, atomic layer step edges, or previously formed defects. Interestingly, areas adjacent the affected regions maintain high structural integrity and crystalline quality as seen in Figure 5b (Figure S11, supporting information). This is in sharp contrast to the rapid heating case where the entire flake/crystal simultaneously disintegrates into nanoscale, faceted and highly crystalline islands of stoichiometric MoS$_2$. Multiple ex-situ experiments have been carried out to determine the required thermal conditions to initiate the disintegration process in MoS$_2$ layers (See Figure S12 for more details). The formation of S-vacancies as the temperature is slowly increased leads to a volatile ejection of sulfur compounds, reducing the composition of sulfur in the lattice. This mechanism has been well explored at atomic level in previous reports[24, 52, 53]. Since there is a net loss of S from the lattice due to its lower energy for vacancy formation, this results in a non-stoichiometric decomposition which leads to the formation of nanocrystalline to amorphous MoS$_2$ or Mo nanocrystals. This can be clearly seen in the atomically resolved HAADF-STEM images as well as the corresponding Fast Fourier transform (FFT) pattern analyses (Figure 5c-d). To further clarify our *ex-situ* observations of samples subjected to a slow heating rate, we also carried out the same thermal decomposition for samples held at fixed temperature (525 ˚C) for different lengths of time, as shown in Figure S13 (supporting information). These HAADF-STEM images provide insight into the mechanistic behaviour during ex-situ thermal decomposition: as the time at 525 ˚C increases, the formation rate and areal density of pores also increases, which leads to increased void formation in the multilayers and monolayers (Figure S12c). After 10 minutes of annealing, bright contrast regions begin to appear, which indicates the formation of $Mo/MoS_x$ nanocrystals along with adjacent regions of amorphized $MoS_2$. This suggests that the non-stoichiometric disintegration first results in formation of nanocrystalline or amorphous $Mo_xS_{2-x}$ followed by near complete removal of S to form Mo metal nanocrystals. Additional atomic resolution images and compositional analysis are provided in supporting information Figures S11-13. To verify that the ex-situ, slowly heated samples in a furnace are not affected by trace oxygen or different quality of vacuum in a tube furnace vs. the TEM column, we have also performed control experiments in an inert, ultra-high purity Ar environment where we observed similar results (see Figure S14). Here, the increased vapor pressure of the Ar environment as compared to vacuum raises the disintegration temperature, consistent with our proposed mechanism described below (Figure 6).

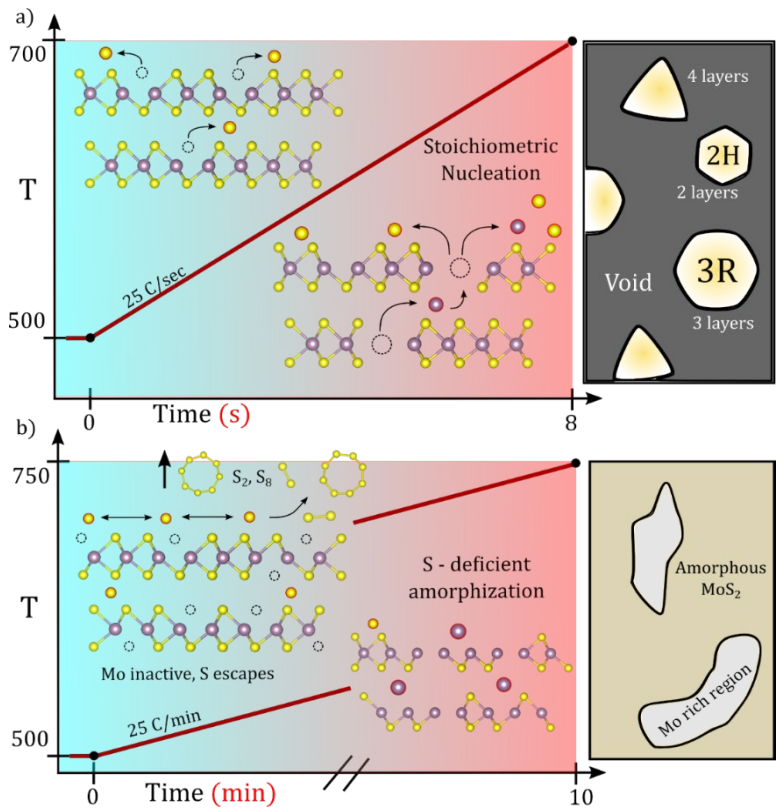

**Figure 6:** (a) Schematic diagram showing that the rapid heating rate leads to the formation of S vacancies and adatoms (yellow circles). Diffusion of S vacancies allows Mo adatoms (purple circles) to recombine with S adatoms before they evolve as volatile elemental products (yellow, $S_2$, $S_8$) and leave the system. (b) In contrast, a slow heating rate prevents Mo adatom formation before S-compounds become volatile, preventing subsequent reformation of Mo-S bonds. This results in the system being driven into a two-phase regime composed of Mo-rich regions and amorphous $MoS_2$ regions.

During rapid in-situ heating, the samples are heated at 25 °C/sec from 500 °C to 700 °C (Figure 6a), while during the ex-situ experiment (S14) shown in Figure 6b the sample was heated over the same temperature range at a rate of 25 °C/min. The non-equilibrium heating rate (Figure 6a) causes sulfur vacancies to form quickly, leading to S-adatoms on the $MoS_2$ surface. Over time, collisions between the rapidly expanding adatom population lead to formation of volatile elemental sulfur compounds ($S2$, $S8$) which leave the surface. However, in only a few seconds, the temperature increases to the point where Mo vacancies begin to form (700 °C), leading to Mo adatoms on the $MoS_2$ surface. This is assisted by the formation of sulfur vacancy line defects, which reduce the formation energy of a Mo vacancy by increasing the local chemical potential of Mo[16, 45]. The chemical potential imbalance occurs because the diffusion rate of the S vacancies is much faster (barrier of 0.8 eV) than that of the S adatoms (barrier of 1.6 eV), which allows the Mo adatoms to begin forming before the volatile S compounds form and leave the surface[45, 54]. The combination of line defects and sulfur conservation on the surface leads to both void formation and a CVD – type environment for stoichiometric nucleation. CVD growth studies have shown that temperatures of 700 ˚C - 800 ˚C preferentially lead to vertically stacked $MoS_2$ growth instead of monolayer growth[47, 54]. Random nucleation of these additional layers can lead to a mixture of 2H and 3R stackings, but as voids become larger and multilayer islands become thicker, we expect a shift toward the 3R morphology below 1000 ˚C, consistent with the bulk phase diagram[16]. In contrast, the sample in Figure 6b spends several minutes in the regime where sulfur vacancies can form but where the temperature is too low to activate Mo adatoms, due to the lower heating rate. The slow heating rate allows for enough time for volatile sulfur compounds to form and leave the surface. This explains the increased depression of the melt process in vacuum versus the neutral Ar environment, where the reduced vapor pressure accelerates the formation and evolution of the volatile sulfur rich compounds. This pushes the remaining atomic composition of the lattice into the $Mo + MoS_2$ or $Mo + Mo_2S_3$ regimes on the bulk phase diagram. The phase transition temperature is affected by both the vapor pressure and the melting point suppression effect of nanoconfinement. At this reduced sulfur concentration, we expect two-phase regions of Mo and either $MoS_2$ or $Mo_2S_3$ to form, as observed in the ex-situ heating experiments.

## Conclusions:

In conclusion, we have performed a detailed analysis of thermal decomposition and disintegration of atomically thin $MoS_2$ via electron microscopy and spectroscopic analysis. We observe that high heating rates lead to spontaneous disintegration of $MoS_2$ single crystalline flakes into highly crystalline, nanosized crystals down to 5 nm in diameter, which is in the regime of strong electronic quantum confinement[32]. Furthermore, we observe that this rapid heating leads to a partial disintegration and the subsequent formation of 3R phases in close proximity to the original 2H phase. The nanocrystals in the partially disintegrated region maintain epitaxial relationship with the underlying layer of $MoS_2$ while the fully disintegrated regions show random orientation of $MoS_2$ nanocrystals. In contrast, slow heating results in non-stoichiometric decomposition of $MoS_2$ single crystals, resulting in first formation of nanocrystalline to amorphous $Mo_xS_{2-x}$ and ultimately Mo metal nanocrystals due to sulfur sublimation. Our study delineates the pathways by which thermally induced disintegration of atomically thin chalcogenides occurs and elucidates the impact of heating rates on the reaction products. This is critical knowledge in applications such as phase change memory, high temperature lubrication, and catalytic processes where $MoS_2$ can be used. In addition, our studies have shown a novel pathway towards creation of an atomically-thin, highly ordered, quantum-dot like structures via top-down processing which could exhibit unique and novel electronic and optical properties for optoelectronic or catalytic applications.

## Experimental Details:

### Materials and Methods:

Mechanically exfoliated 2D $MoS_2$ layers were prepared using the conventional scotch tape method, as described elsewhere[55]. The thickness of the exfoliated layers has been intentionally chosen such that each sample could be reproduced for further thermal diffusion analysis. Exfoliated $MoS_2$ layers were transferred to the TEM chips/grids with help of a poly-dimethyl siloxane (PDMS) strip, using a dry transfer technique that utilizes micromanipulator stages attached to an optical microscope. After transferring $MoS_2$ to TEM grids, samples were annealed in a quartz tube furnace in a closed gas (Ar+ $H_2$) environment to remove all PDMS contaminants. Annealing was performed at 300 ˚C for 2 hrs to clean contaminants as well as release the strain developed during the pressure-based transfer method.

### STEM Measurements:

Scanning transmission electron microscopy (STEM) has been used to measure and analyse all the samples made for in-situ and ex-situ heating experiments. In-situ experiments at various temperatures have been done in a specialised TEM holder from Hummingbird Scientific, where a uniquely designed microfabricated heating chip was used. An embedded heating element provides robust stability to perform in-situ thermal diffusion analysis at elevated temperatures. Heating calibration was received from the manufacturer to achieve controlled heating based on resistance measurements. Two different TEM/STEM systems were used: a JEOL F200 S/TEM, and probe-corrected JEOL NEOARM STEM, both operated at 200 KV accelerating voltage. For the JEOL NEOARM STEM, the condenser lens aperture was 40 μm with a camera length of 4 mm for imaging and 2 mm for EELS. The probe current was 120 pA. A Gatan imaging filter (GIF) aperture of 2.5 mm was also used for EELS with a collection semi-angle β of 41.7 mrad. All of the captured TEM images were collected on GATAN IS One View (JEOL F200) and Ultra-scan cameras (JEOL NEO ARM), and the STEM images were recorded on the integrated JEOL bright-field and annular dark field detectors. In-situ TEM videos were acquired on the JEOL F200 using GATAN OneView IS camera at 50 fps with 2k x 2k resolution. Corresponding elemental identification has been performed using dual detector EDX on the F200 system. Experimentally acquired STEM images are smoothed using the adaptive gaussian blur function (with radius of 1-2 pixels) available in ImageJ. Image simulation has been performed using QSTEM software[34]. Adopted parameters for the image simulation were the same as in the experimental conditions for AC-STEM measurements. Image analysis and feature extraction were preformed using the open source SciKit-Image[46] python library.

## Acknowledgements:

This work was carried out in part at the Singh Center for Nanotechnology at the University of Pennsylvania which is supported by the National Science Foundation (NSF) National Nanotechnology Coordinated Infrastructure Program grant NNCI-1542153. D.J., E.A.S and P.K acknowledge primary support for this work from University of Pennsylvania Materials Research Science and Engineering Center (MRSEC) seed grant supported by the National Science Foundation (DMR-1720530) and NSF DMR Electronic Photonic and Magnetic Materials (EPM) core program grant (DMR-1905853). N.A. and D.J. acknowledge support from Vagelos Institute for Energy Science and Technology at University of Pennsylvania as well as the Center for Undergraduate Research and Fellowships at Penn. D.J. also acknowledges support for this work by the U.S. Army Research Office under contract number W911NF1910109. J.P.H and E.A.S acknowledge support through the National Science Foundation, Division of Materials Research, Metals and Metallic Nanostructures Program under Grant 1809398. A.C.F and E.A.S would like to acknowledge the Vagelos Institute for Energy Science and Technology at the University of Pennsylvania for a graduate fellowship. C.C.P. and V.B.S acknowledge support from NSF grant CMMI-1727717. The authors thank Douglas Yates and Jamie Ford in the Singh Center for Nanotechnology for help with the TEM/STEM measurements.

# Supporting Information

# Direct Visualisation of Out-of-Equilibrium Structural Transformations in Atomically-Thin Chalcogenides

**Pawan Kumar**[1,2], **James P. Horwath**[2], **Alexandre C. Foucher**[2], **Christopher C. Price**[2], **Natalia Acero**[3], **Vivek B. Shenoy**[2], **Eric A. Stach**[2#] **and Deep Jariwala**[1#].

[1]Department of Electrical and Systems Engineering, University of Pennsylvania, Philadelphia-19104, USA.

[2]Department of Materials Science and Engineering, University of Pennsylvania, Philadelphia-19104, USA.

[3]Vagelos Institute for Energy Science and Technology, University of Pennsylvania, Philadelphia-19104, USA.

**#E-mail: dmj@seas.upenn.edu; stach@seas.upenn.edu**

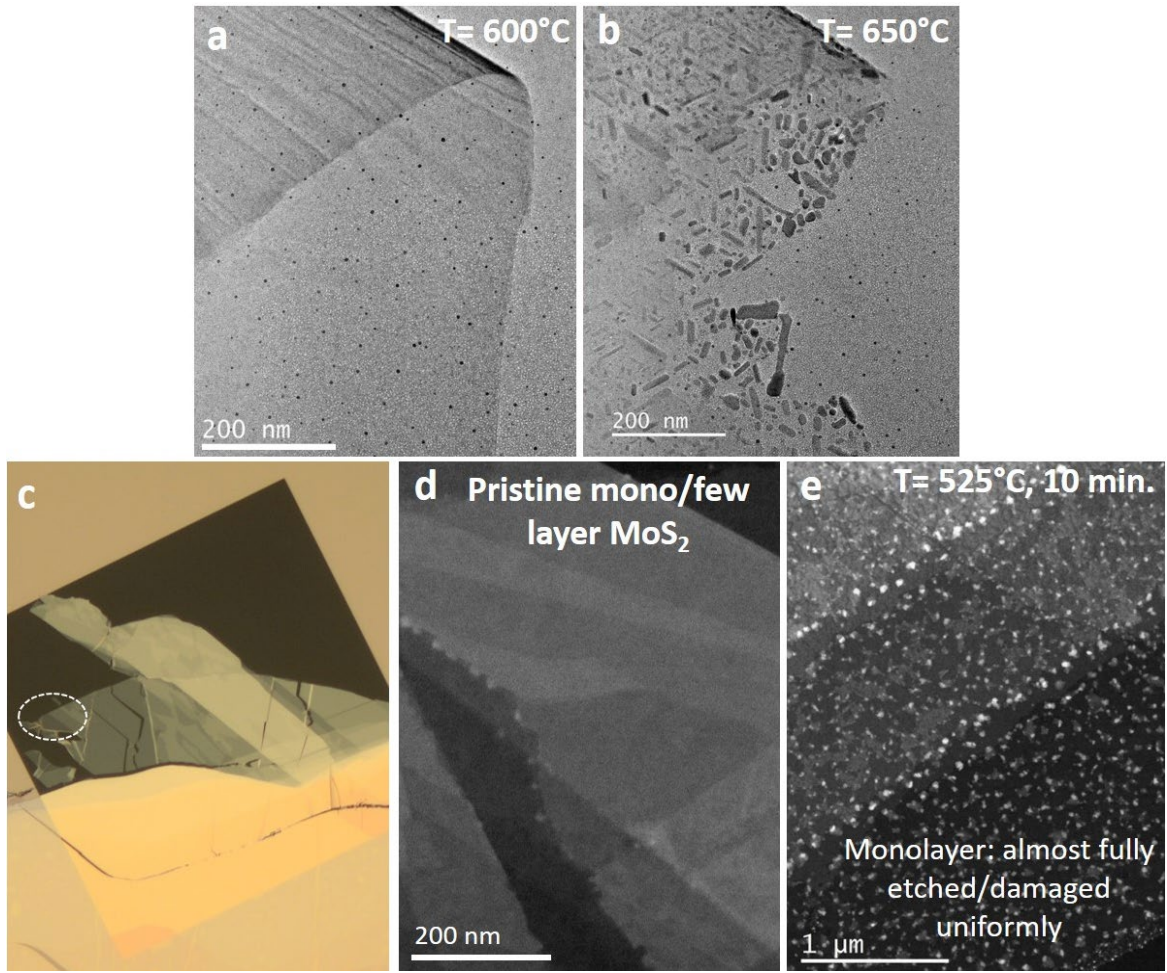

**Figure S1: Non-equilibrium vs. equilibrium thermal diffusion processes within 2D MoS₂ layers.** (a) ADF STEM images during in-situ heating of MoS₂ layer indicate no changes to the sample at temperatures of 600 ºC, while (b) a sudden (25 ºC/sec) temperature increase to 650 ºC caused crystalline disintegration of 2D layers into faceted nanostructures. (c) Optical view of SiNx membrane

coated ex-situ heating TEM grid and corresponding ADF STEM image of (d) as prepared few layer MoS$_2$ and (e) after ex-situ heating (heat ramping rate 25 ºC/min) at 525 ºC for 10 minutes under a rough vacuum environment.

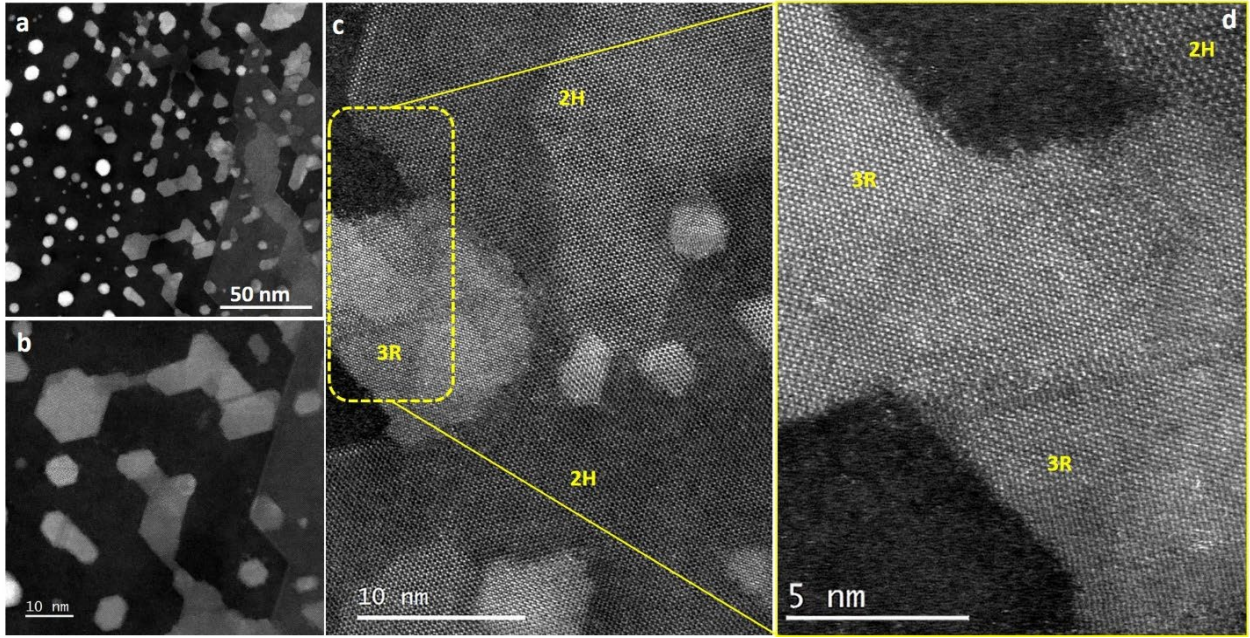

**Figure S2: In-situ heated bilayer MoS$_2$ in a non-equilibrium environment and corresponding magnified HAADF-STEM imaging of polymorphic-phases.** (a, b) Low magnified dark field STEM images after in-situ disintegration at 700 ºC in non-equilibrium conditions (25 ºC/sec) and corresponding (c) atomic resolution HAADF STEM image. (d) Highlighted portion in Figure c, shows restructured polymorphic phases, 3R (different thickness) from the pristine 2H phase of bilayer MoS$_2$.

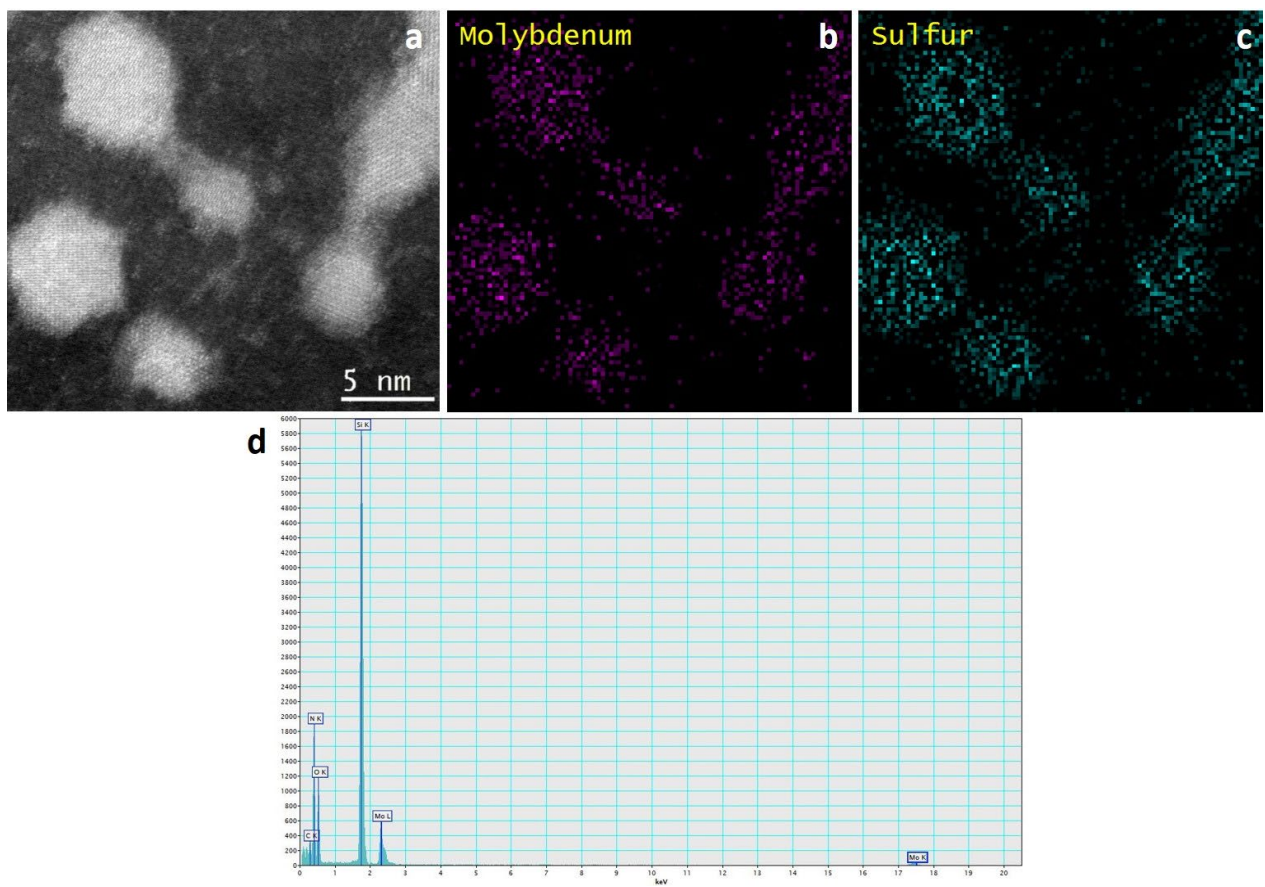

**Figure S3: EDS maps and spectrum of the crystalline disintegrated MoS₂.** (a) HAADF STEM image and corresponding elemental identification (b) Mo and (c) S, the full spectrum is presented in (d).

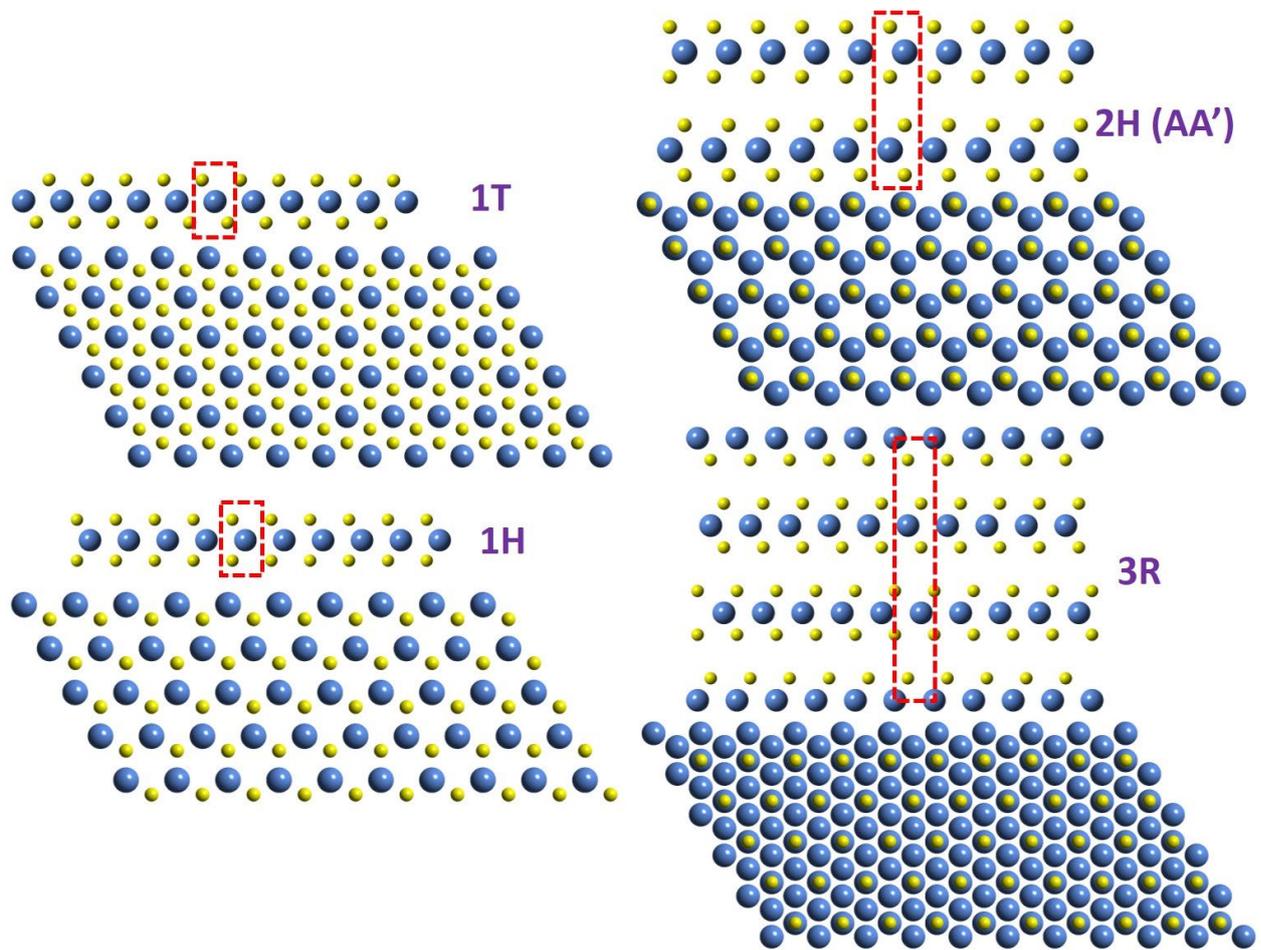

**Figure S4: Structural models:** Structural model (along a and c axes) with marked unit cell representation extracted using Crystal Maker for different polymorphic phases of $MoS_2$ according to the respective lattice parameters and space-group. This structural model allows us to directly determine and visualize the polymorphic phases in the atomic resolution HAADF-STEM images.

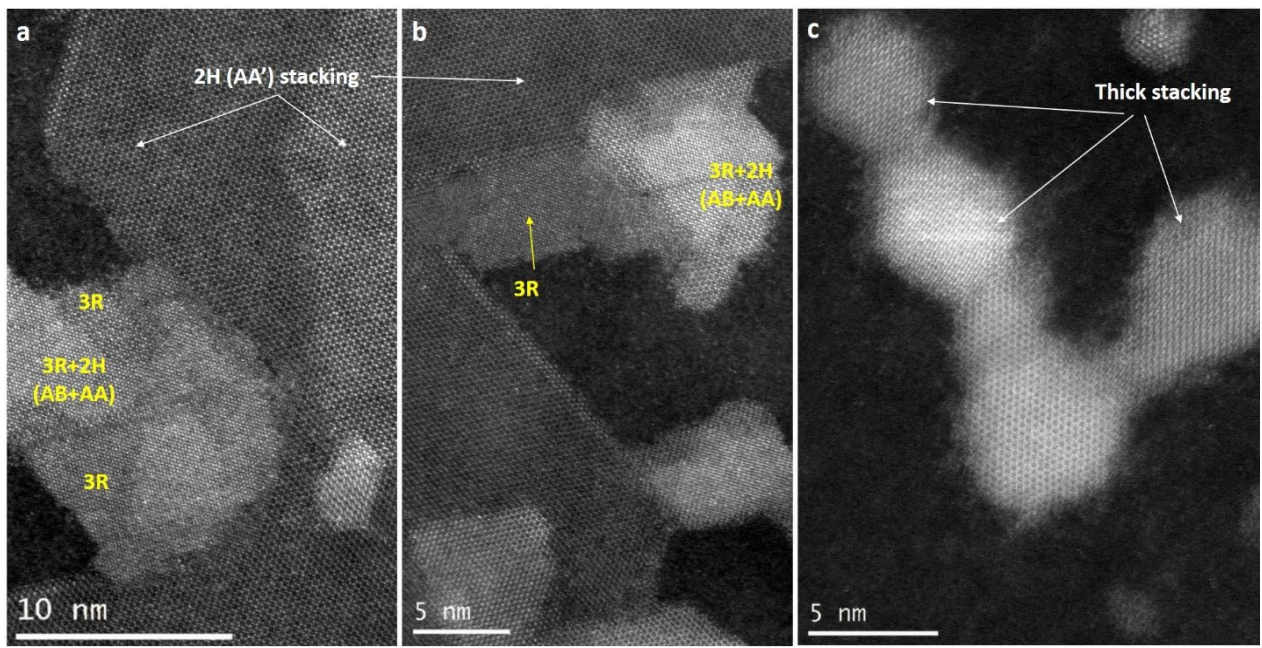

**Figure S5: Atomic resolution STEM imaging of the highlighted portion of disintegrated bilayer MoS$_2$ having different stacking sequences.** a-b shows the atomically resolved structural arrangement of two different phases. Interestingly, we can see that the rearrangement and formation of different phases only happens when the nanostructure is completely disintegrated. Here, bilayer MoS$_2$ remains in the original 2H arrangement when it is in a continuous layer

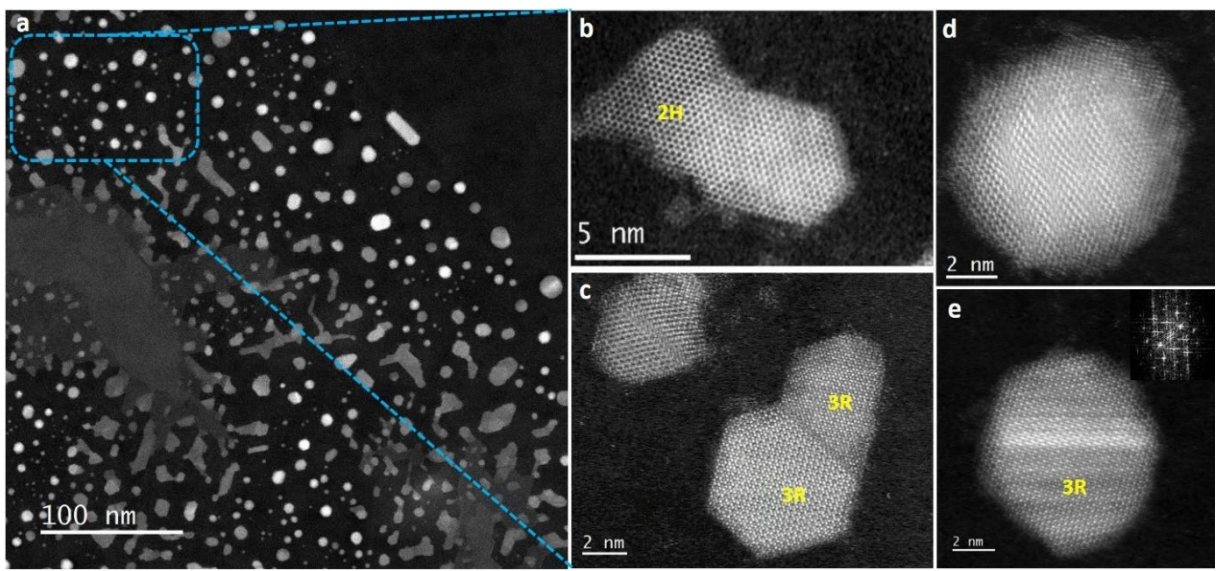

**Figure S6: STEM imaging of disintegrated quantum dot like particles.** (a) Atomic resolution ADF-STEM imaging for the in-situ disintegrated bilayer $MoS_2$ and higher magnification of individual quantum dots (b-e).

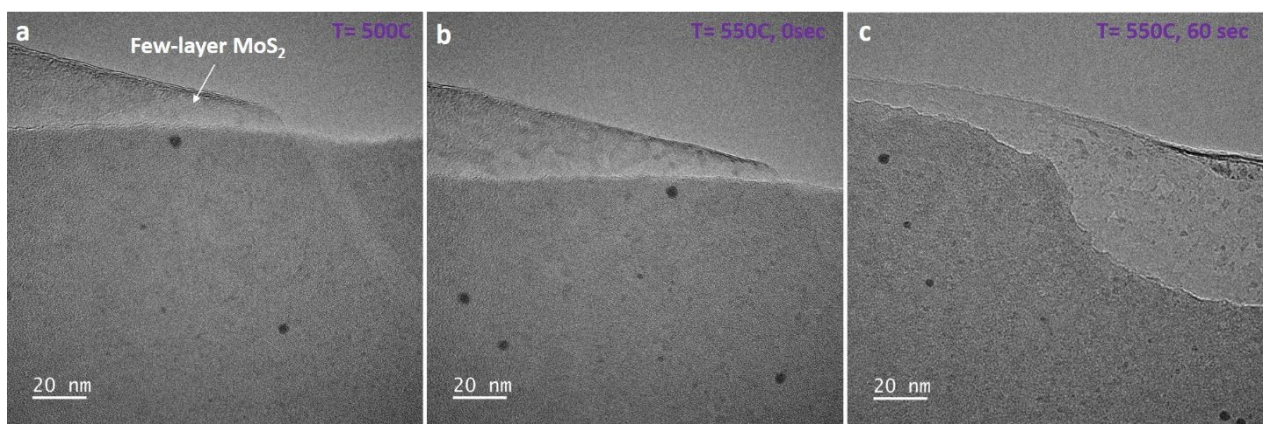

**Figure S7: Time sequence TEM images acquired during in-situ heating of $MoS_2$ from 500 to 550 ºC at a slow heating rate (25 ºC/min).**

**At T= 650°C with increasing time (1-6)**

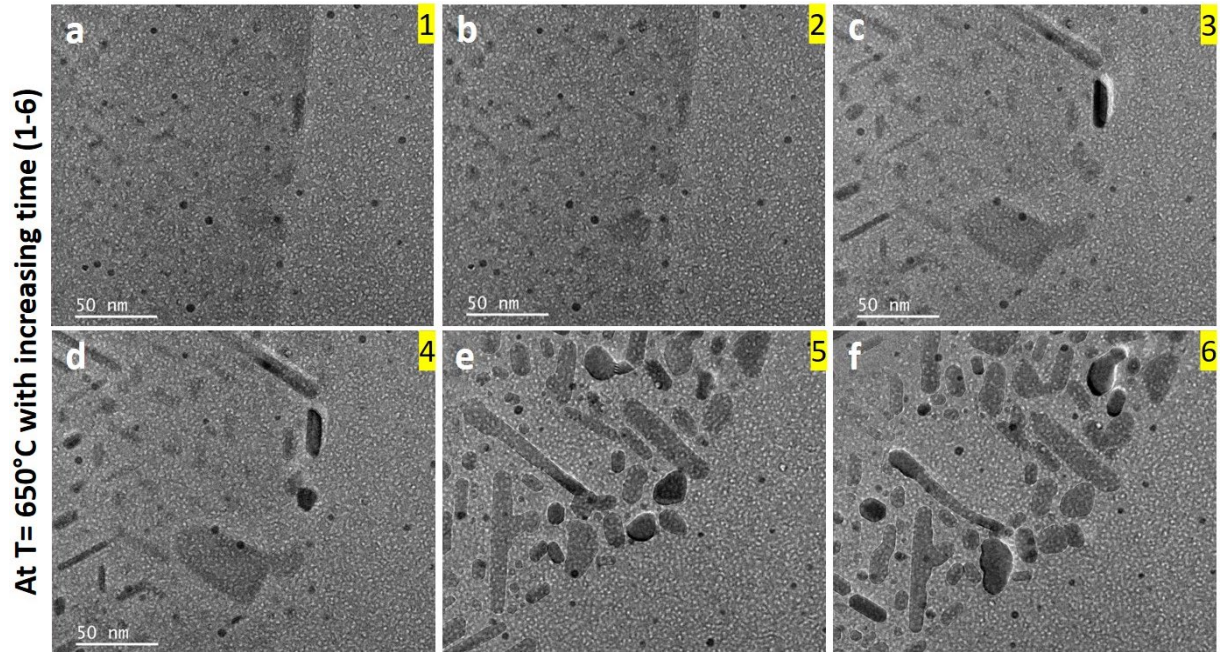

**Figure S8: Sequential TEM images captured following rapid sample degradation.**

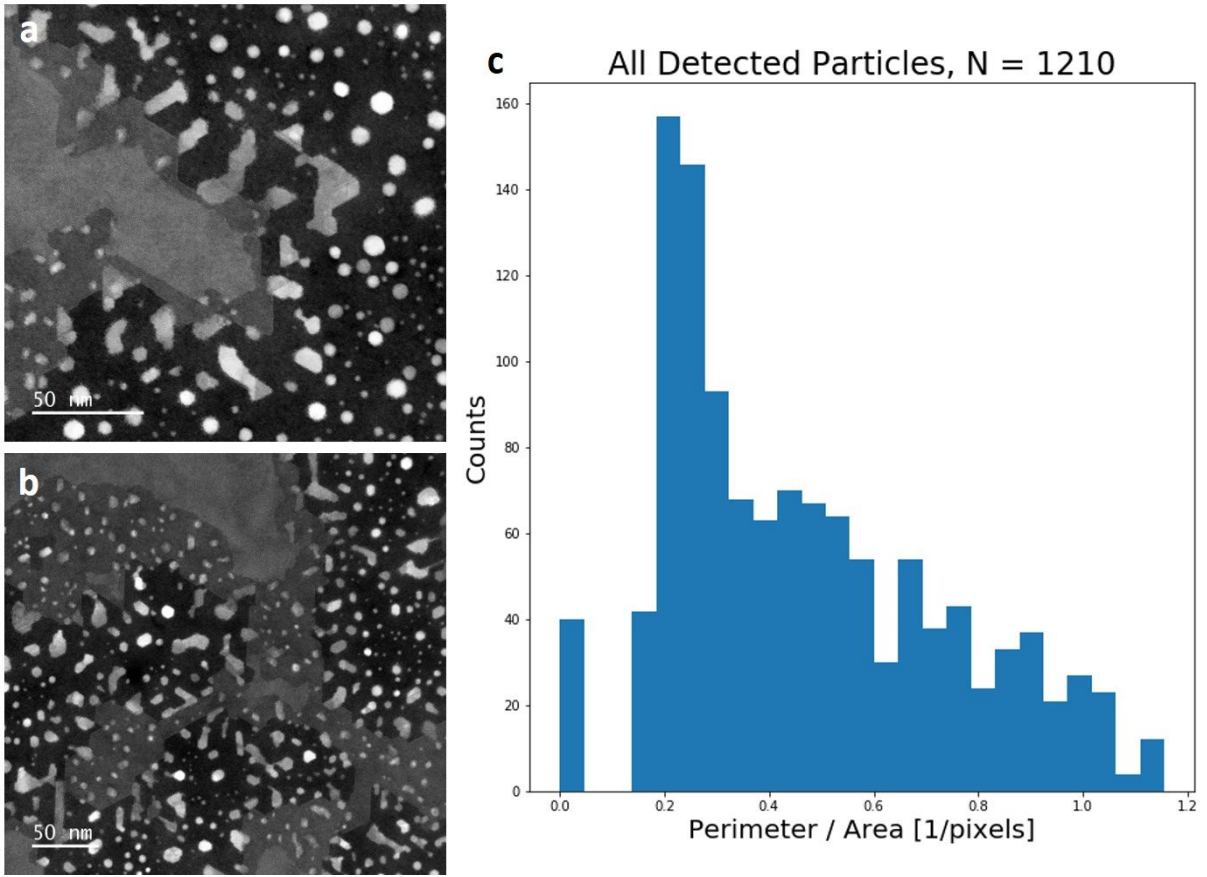

**Figure S9: Shape and size analysis of well disintegrated MoS₂ using python-programmed image pixel-contrast relationship.**

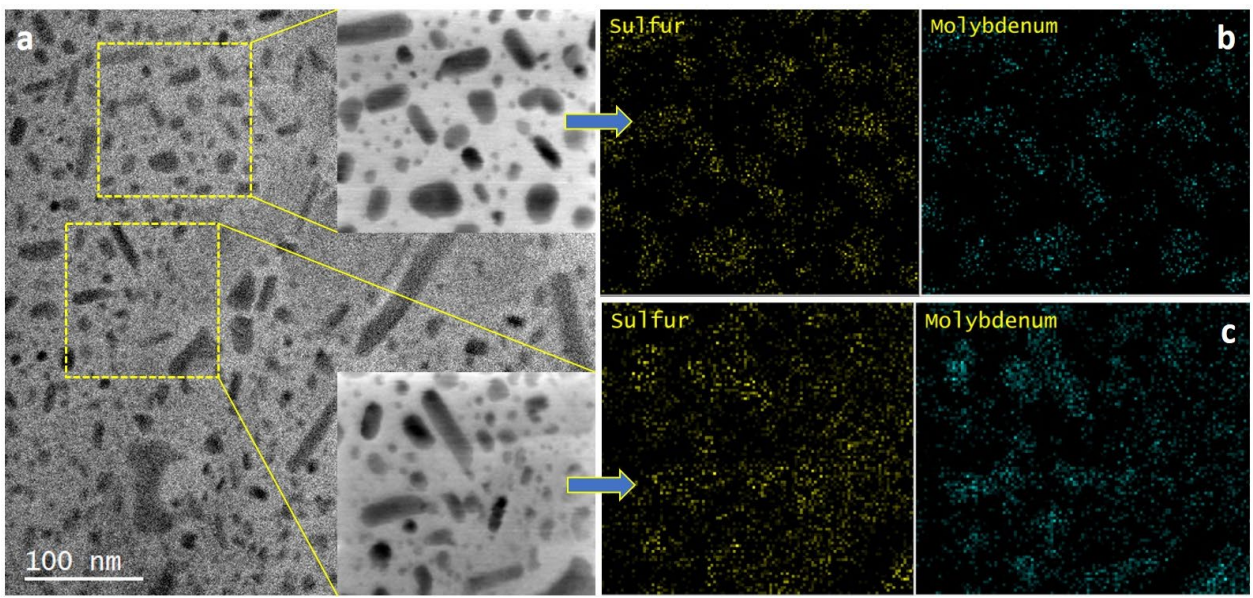

**Figure S10: Additional EDS maps for a different sample heated in-situ, confirming the disintegrated nanostructured MoS₂.**

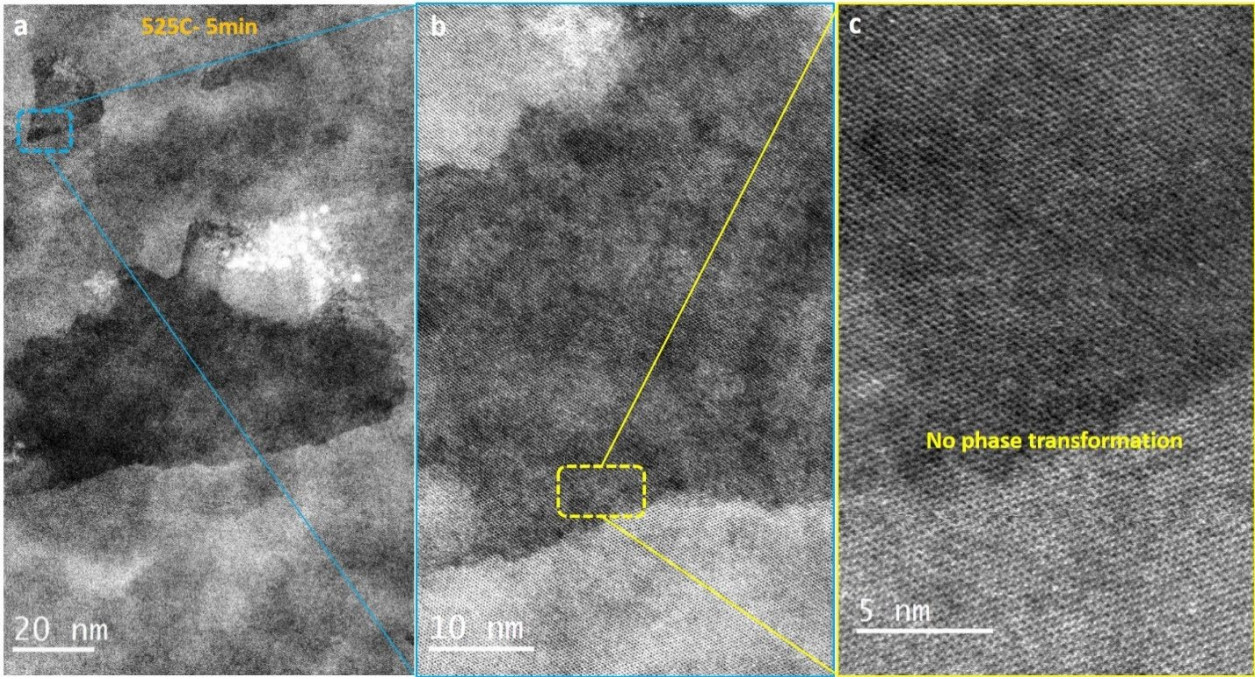

**Figure S11: Atomic resolution STEM images of bilayer MoS₂, heated ex-situ to 525 ºC for 5 minutes.**

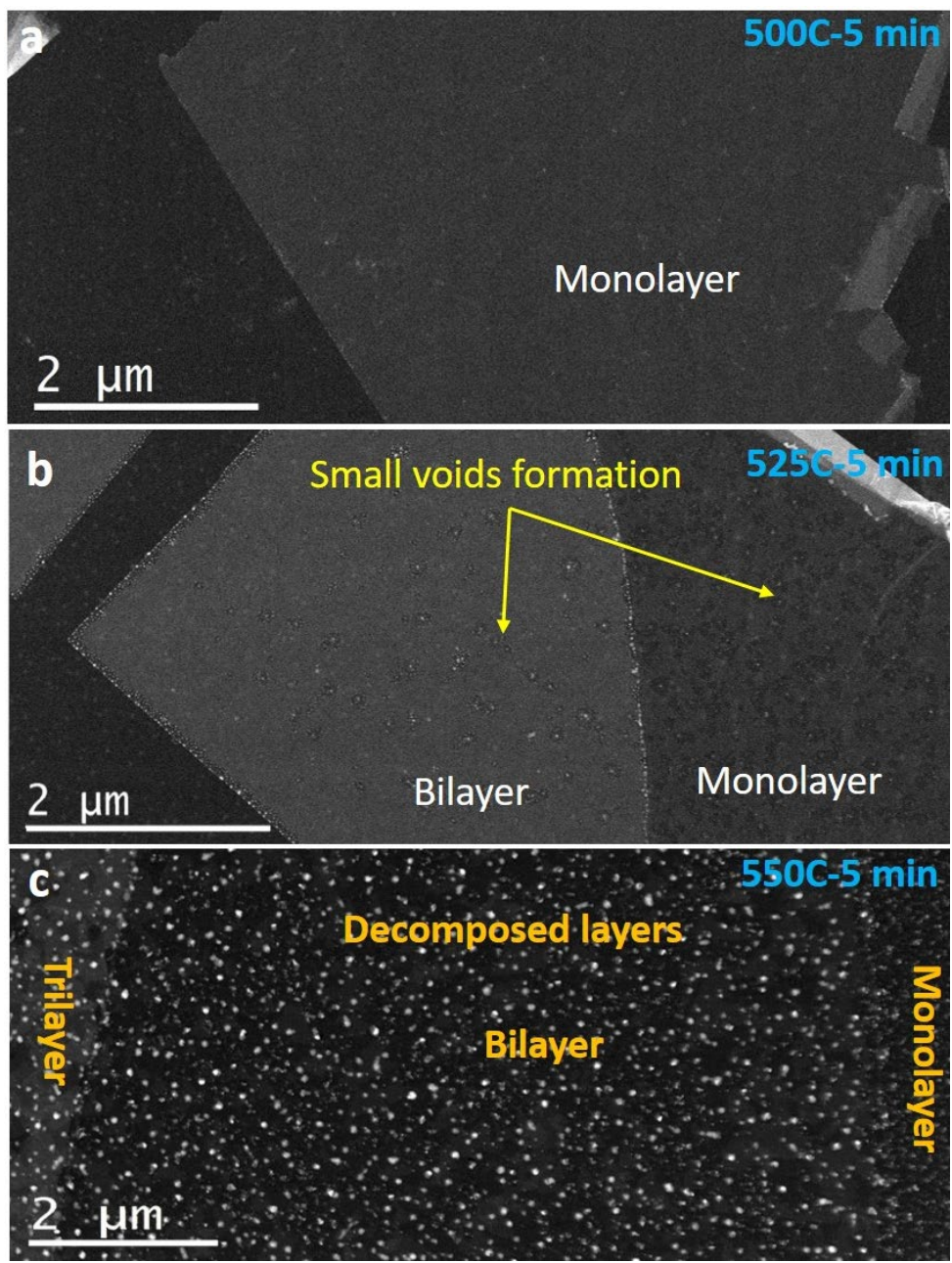

**Figure S12: HAADF-STEM images of MoS$_2$ at different temperatures (rate = 25 ºC/min) to analyse the thermal diffusion behaviour.**

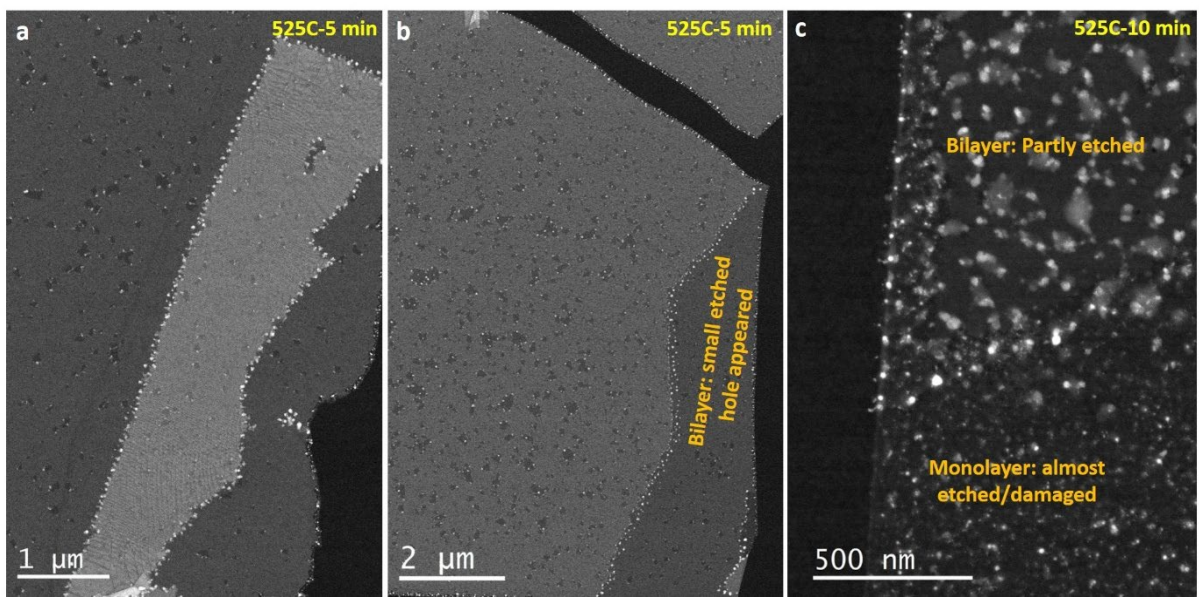

**Figure S13: Set of ADF-STEM images highlight the different regions of MoS₂ heated ex-situ (Rate = 25 ºC/min) at 525 ºC for two different times.**

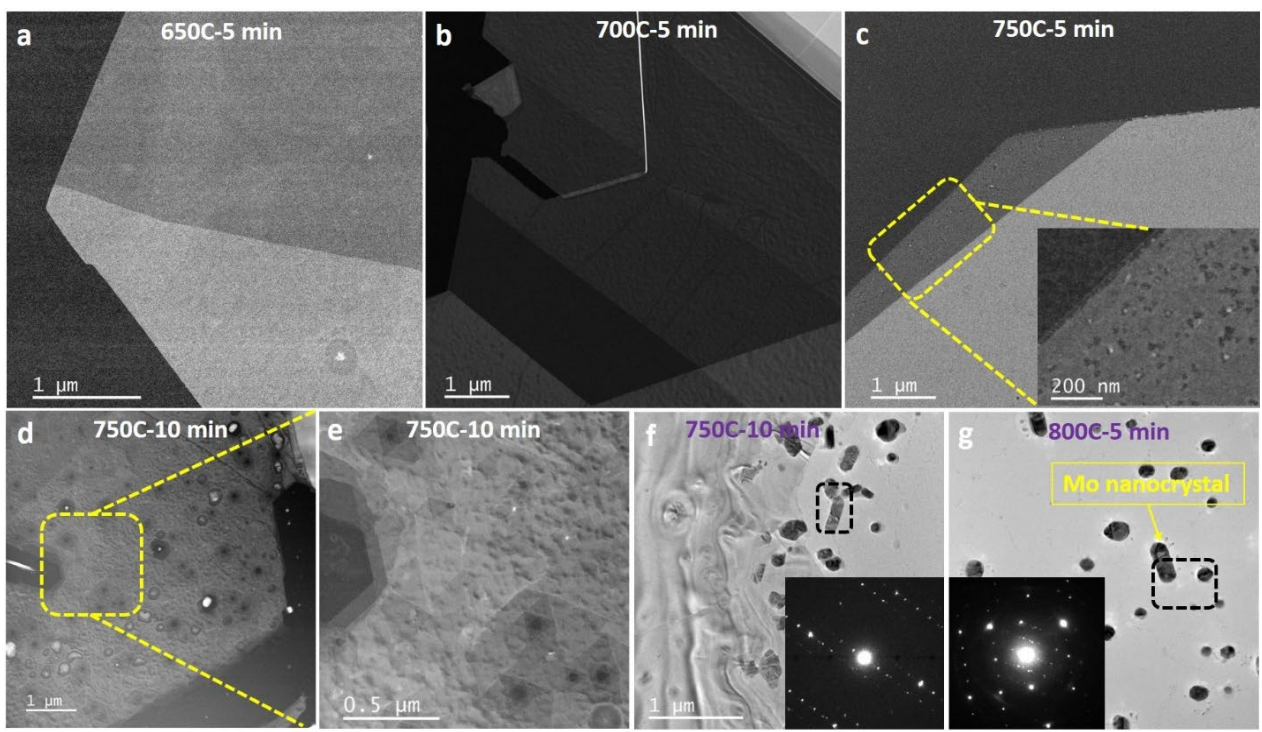

**Figure S14: HAADF-STEM images from few-layer MoS₂ heated at the slow ramping rate (Rate = 25 ºC/min) in an argon (Ar) environment under equilibrium conditions.** A slower etching rate and further decomposition occurs at higher temperatures in the Ar environment when compared with the rough vacuum. HAADF-STEM images of ex-situ heated MoS₂ layers at (a) 650 ºC, (b) 700 ºC, (c) 750 ºC for 5 minutes. Figure c inset shows magnified region of smaller etching in thinner section of figure c. (d) STEM image following 10 minutes at 750 ºC. This leads faceted (triangular or hexagonal shaped) features shown at higher magnification shown in figure (e). A few portions of the MoS₂ layers become complete decomposed into Mo nanocrystals as shown in (f), which is further confirmed by another sample 800 ºC for 5 minutes. The SAED pattern confirms the transformation into Mo nanocrystals.